# Mode localization phenomenon in microbeams due to surface roughness

**Mohamed Shaat**∗

*Engineering and Manufacturing Technologies Department, DACC, New Mexico State University, Las Cruces, NM 88003, USA*

*Mechanical Engineering Department, Zagazig University, Zagazig 44511, Egypt*

**Abstract**

This is the first study on the mode localization phenomenon in microbeams due to surface roughness. A new model for microbeams with rough surfaces is developed. The natural frequencies and mode shapes of cantilever, simple supported, and clamped-clamped microbeams are determined depending on the beam surface roughness. A parametric study is presented demonstrating two prospects: surface roughness may lead to a zero-frequency mode or a mode localization. As for the first prospect, it is demonstrated that surface roughness may add more softness to a specific mode of vibration and reduce its natural frequency causing a rigid-body mode. As for the second prospect, surface roughness may inhibit the propagation of vibration energy throughout the beam length leading to a mode localization. It is revealed that a mode localization is accompanied with an increase in the natural frequency of the microbeam. It is revealed that the description of the beam vibration according to one of these two prospects depends on the beam size, parameters of the surface roughness, and the boundary conditions.

**Keywords**: mode localization; zero-frequency mode, roughness; microbeams; vibration.

## 1. Introduction

Micro/nano-sized structures are intensively used to design advanced micro/nano-devices for various engineering and medical applications. Because of their unique characteristics and the promises that they can give, micro/nano-structures are implemented to achieve optimized performance of aerospace, automotive, medical, and defense applications. These small-scale structures allow engineers control their microstructures, surfaces, and sizes to develop designs of smart and multi-functional systems.

Conventional structural analysis methods assume ideal structures (free of irregularities). However, due to uncontrollable manufacturing-based errors, engineering structures usually possess irregularities in the form of material and/or geometrical variations. The presence of these irregularities in a structure may result in drastic changes in its dynamical behaviors. For instance, due to a small irregularity in a nearly periodic structure, the vibration is inhibited and localized in a specific region of the structure near the source of vibration [Kim and Lee, 1998; Chan and Liu, 2000; Hodges, 1982; Bendiksen, 1987; Pierre, 1988; Cai and Lin, 1991; Triantafyllou and Triantafyllou, 1991; Natsiavas, 1993; Xie, 1995; Cai et al., 1995; Liu et al., 1995; Liu et al., 1996; Pierre et al., 1996]. This phenomena is known as 'Mode Localization'. Mode localization is one of the common changes in the normal modes of structures due to involved irregularities. Mode localization is the confinement of vibration; thus, the magnitude of vibration in a specific region of the structure is large relative to the magnitude of vibration in the rest of the structure. In solid state physics, the localization of electron fields in disordered solids was first observed by Anderson [1958]. In turbomachinery, mode localization was revealed to explain the unpredictable fatigue failure of mistuned blades of turbomachines [El-Bayoumy and Srinvasan, 1975]. In structural dynamics, Hodges [1982] revealed the wave localization in periodic structures with disorders. Afterwards, Bendiksen [1987] investigated mode localization in cyclically-symmetric large space-structures using analytical and numerical methods. Utilizing a perturbation approach, Pierre and Dowell [1987] explained the localization of modes of two coupled-pendulums with system disorders. Pierre et al. [1987] experimentally revealed the mode localization of a weakly coupled two-span beam with a small irregularity. They used the modified perturbation method to explain the influences of the structural disorder and coupling strength on the mode localization. Lust et al. [1993] revealed mode localization in multispan-Timoshenko beams.

---

∗Corresponding author. Tel.: +15756215929.
 *E-mail addresses:* shaat@nmsu.edu; shaatscience@yahoo.com (M. Shaat).

In addition to mode localization, two successive frequencies of a structure may approach each other and then veer apart with a high local curvature [Chan and Liu, 2000; Pierre, 1988; Liu et al., 1995; Sari et al., 2017]. This kind of behavior is known as 'Eigenvalue Loci Veering'. The swerving of frequencies is usually accompanied with a rapid variation in the eigenvectors which results in a mode shape change [Sari et al. 2017] or mode localization [Pierre, 1988]. Leissa [1974] was the first to discover eigenvalue loci veering in periodic structures. Afterwards, Pierre [1988] investigated the mode localization and the eigenvalue loci veering phenomena in nearly periodic structures with small irregularities. The frequency and mode veering phenomena were discovered in pressure vessels [Doll and Mote, 1976], cables [Triantafyllou 1984; Behbahani-Nejad and Perkin, 1996], stressed frame structures [Du Bois et al., 2009], lattice dynamics [Anderson, 1958; Montroll and Pots, 1955; Rosenstock and Mcgill 1962], and structural dynamics [Hodges, 1982; Valero and Bendiksen, 1986; Xie, 1995; Cai et al., 1995; Pierre et al., 1996].

Studies were conducted to reveal the eigenvalue loci veering and mode localization phenomena in continuous structures. Li et al. [2005] revealed the mode localization of an axially compressed-rib stiffened rectangular simple supported plate. Chen and Xie [2005] investigated the localization of modes of the free and forced vibrations of rib stiffened-rectangular plates. The mode localization of thin clamped Kirchhoff rectangular plates was studied by Filoche and Mayboroda [2009]. Sharma et al. [2012] investigated the mode localization in composite laminates. Verma et al. [2014] revealed mode localization in single and multi-layer graphene nanoribbons. Paik et al. [2015] investigated the localization of buckling modes in plates and laminates.

Although a great deal of progress has been made in revealing eigenvalue loci and mode veering and mode localization phenomena in large-scale structures, the literature lacks of studies on these phenomena in micro/nano-scale structures. The frequency and mode veering phenomena in micro/nanobeam-structures with material and/or geometrical irregularities were revealed in a few studies. Dehrouyeh-Semnani et al. [2016] explored the mode veering phenomenon in curved functionally graded microbeams. Sari et al. [2017] revealed frequency and mode veering phenomena in axially-functionally graded non-uniform beams. They demonstrated that due to nonlocal residuals and the material and geometrical irregularities, two successive frequencies may approach each other and veer away resulting in a mode change. Pradiptya and Ouakad [2017] demonstrated that when a buckled nanobeam is exposed to high temperature, mode veering may take place. Indeed, micro/nano-structures are more sensitive to material and geometrical irregularities than large-scale structures. Therefore, more studies on exploring the changes in the dynamical characteristics of nanostructures due to small irregularities should be conducted. Exploring the eigenvalue loci and mode localization in micro/nano-structures is of a particular interest where MEMS resonators for sensing applications utilizing mode localization are being designed [Zhao et al., 2016].

To fill in the aforementioned missing fold, the current study presents the first model of beams with rough surfaces exploring the mode localization phenomenon in microbeams due to the roughness of their surfaces. The model presented in this study depends on Shaat's surface integrity model. This model explores surface integrity effects (*e.g.* surface roughness, surface waviness, altered layers, and surface excess energy) on the mechanics of nanomaterials, *e.g.* ultra-thin films [Shaat, 2017]. Shaat's surface integrity model outweighs Gurtin-Murdoch surface elasticity model for accounting for the effects of coupling the surface excess energy with the surface roughness and waviness. Thus, the former model assumes smooth surfaces of materials. However, a real surface of a material is textured with many forms of irregularities, *e.g.* roughness, waviness, cracks, etc. In this study, effects of the surface irregularities in terms of surface roughness on the frequencies and mode shapes of microbeams with various boundary conditions are explored.

This effort presents the first study on the mode localization phenomenon in microbeams due to surface roughness. A formulation of surface roughness effects on the mechanics of microbeams is presented in Section 2. The equation of motion is derived containing measures for the surface roughness. Then, the derived model is solved for the free vibration of cantilever, simple supported, and clamped-clamped beams in Section 3. The natural frequencies along with their corresponding mode shapes are determined. In Section 4, the influence of the surface roughness on the natural frequencies and mode shapes of microbeams are discussed. Moreover, the mode localization phenomena in microbeams due to surface roughness are explored. The findings of this study provide new insights on the influence of surface roughness on the mechanics of nanostructures.

## 2. Modeling beams with rough surfaces

Consider a microbeam consists of a material bulk with a volume $V$ surrounded by a rough surface $S$, as shown in Figure 1. To account for the surface roughness effects, $\mathcal{R}(x)$ is introduced as the profile of the surface roughness. Based on Euler-Bernoulli beam model, the displacement field, $u_i$, of a point, $x = (x, y, z)$, is defined as follows:

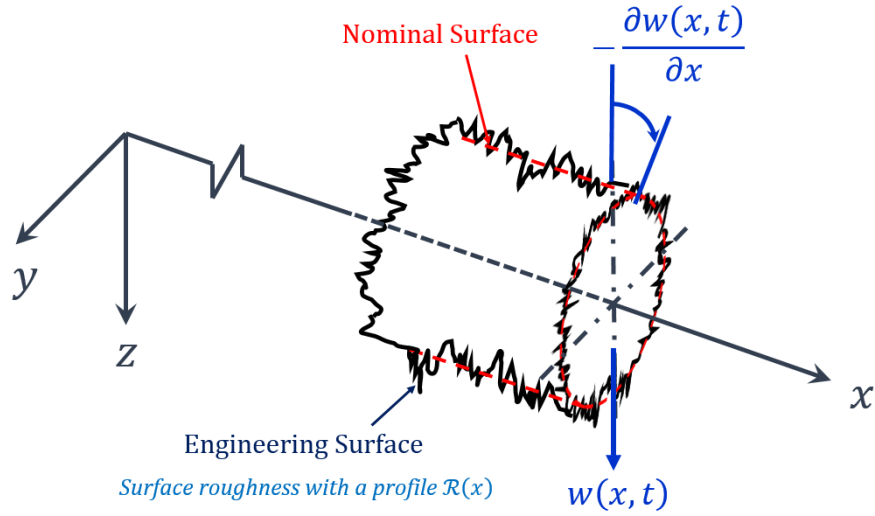

Figure 1: A schematic of an Euler-Bernoulli beam with a rough surface.

$$u_x(x, z, t) = -z \frac{\partial w(x,t)}{\partial x}, \quad u_y = 0, \quad u_z(x, t) = w(x, t) \tag{1}$$

where $w(x, t)$ is the beam deflection.

According to equation (1), the non-zero strain component is obtained as follows (assuming linear elasticity):

$$\varepsilon_{xx}(x, z, t) = -z \frac{\partial^2 w(x, t)}{\partial x^2} \tag{2}$$

Hence, for an isotropic-linear elastic beam, the constitutive equation is defined as follows (neglecting Poisson's ratio effects):

$$\sigma_{xx}(x, z, t) = E \varepsilon_{xx}(x, z, t) \tag{3}$$

where $\sigma_{xx}$ is the beam axial stress. $E$ denotes the beam elastic modulus.

In the present study, the equation of motion of microbeams with rough surfaces are derived based on Hamilton's principle:

$$\int_0^t (\delta T - \delta U + \delta Q) \, dt = 0 \tag{4}$$

where $\delta U$, $\delta T$, and $\delta Q$ are, respectively, the first variations of the strain energy, kinetic energy, and work done which can be defined for the considered beam model as follows:

$$\delta T = -\int_V (\rho \ddot{w} \delta w) dV \tag{5}$$

$$\delta U = \int_V (\sigma_{xx} \delta \varepsilon_{xx}) dV \tag{6}$$

$$\delta Q = \int_V (f_z(x,t) \delta u_z) dV + \int_\Gamma (t_z \delta u_z) dS \tag{7}$$

where $\rho$ is the mass density of the beam material. $f_z(x,t)$ denotes the body force in z-direction. $t_z$ is the surface traction at the beam end, $\Gamma$.

By substituting equations (5)-(7) into equation (4), the equation of motion of microbeams with rough surfaces can be obtained in terms of the stress resultants as follows:

$$\frac{\partial^2}{\partial x^2} M_{xx}(x,t) + F_z(x,t) - I_0(x) \ddot{w}(x,t) = 0 \tag{8}$$

where

$$F_z(x,t) = \int_{A(x)} f_z(x,t) dA$$

$$I_0(x) = \int_{A(x)} \rho \, dA \tag{9}$$

with the corresponding boundary conditions at an end, $\Gamma$:

$$\frac{\partial}{\partial x} M_{xx} = \bar{V} \text{ or } w = w^\Gamma$$
$$M_{xx} = \bar{M} \text{ or } w_{,x} = w^\Gamma_{,x} \tag{10}$$

where $\bar{V}$ and $\bar{M}$ are the applied shear force and bending moment at the end of the beam, $\Gamma$. $w^\Gamma$ and $w^\Gamma_{,x}$ are prescribed deflection and slope.

It should be mentioned that, because of the surface roughness, the beam is non-uniform where its cross sectional area, $A(x)$, varies trough the beam length according to a random function defines the beam surface roughness, $\mathcal{R}(x)$. Accordingly, the moment stress resultant, $M_{xx}$, introduced in equations (8) can be defined for an isotropic-linear elastic beam as follows:

$$M_{xx} = \int_{A(x)}^{0} z\sigma_{xx} \, dA = -D(x)\frac{\partial^2 w(x,t)}{\partial x^2} \tag{11}$$

where the bending stiffness, $D(x)$, is defined as follows:

$$D(x) = EI(x) \tag{12}$$

where

$$I(x) = \int_{A(x)} z^2 \, dA \tag{13}$$

is the area moment of inertia of the beam.

Assuming a rectangular beam with identical surfaces, the area moment of inertia in equation (13) can be explicitly written as follows:

$$I(x) = \frac{4}{3}\left(\frac{b}{2} + \mathcal{R}(x)\right)\left(\frac{h}{2} + \mathcal{R}(x)\right)^3 \tag{14}$$

and the mass per unit length, $I_0(x)$ (introduced in equation (8)), can be written as follows:

$$I_0(x) = 2\rho\left(\frac{b}{2} + \mathcal{R}(x)\right)\left(\frac{h}{2} + \mathcal{R}(x)\right) \tag{15}$$

where $b$ and $h$, denote, respectively, the beam width and thickness.

By substituting equation (11) into equations (8) and (10), the beam equation of motion and the boundary conditions accounting for the roughness of its surface can be written in terms of the deflection as follows:

$$D(x)\frac{\partial^4 w(x,t)}{\partial x^4} + K(x)\frac{\partial^3 w(x,t)}{\partial x^3} + B(x)\frac{\partial^2 w(x,t)}{\partial x^2} - F_z(x,t) + I_0(x)\ddot{w}(x,t) = 0 \tag{16}$$

where

$$\begin{aligned} K(x) &= 2\frac{\partial D(x)}{\partial x} \\ B(x) &= \frac{\partial^2 D(x)}{\partial x^2} \end{aligned} \tag{17}$$

and the boundary conditions:

$$\begin{aligned} -D(x)\frac{\partial^3 w(x,t)}{\partial x^3} - \frac{1}{2}K(x)\frac{\partial^2 w(x,t)}{\partial x^2} &= \bar{V} \text{ or } w = w^\Gamma \\ -D(x)\frac{\partial^2 w(x,t)}{\partial x^2} &= \bar{M} \text{ or } w_{,x} = w^\Gamma_{,x} \end{aligned} \tag{18}$$

The equation of motion (equation (16)) and the boundary conditions (equation (18)) are derived incorporating a measure, $\mathcal{R}(x)$, to account for the surface roughness effects on the mechanics of microbeams.

As demonstrated by the author in [Shaat, 2017], it is challenging to derive analytical solutions for the obtained equation of motion (equation (16)) in its current form. Moreover, the experiment usually reports the average of the surface roughness. Therefore, following the approach proposed by Shaat [2017], the average of the surface roughness and the average slope of the surface roughness are employed:

$$R_a = \langle \mathcal{R}(x) \rangle = \frac{1}{L} \int_0^L |\mathcal{R}(x)| \, dx$$

$$R_S = \langle d\mathcal{R}(x)/dx \rangle = \frac{1}{L} \int_0^L \left| \frac{d\mathcal{R}(x)}{dx} \right| dx \quad (19)$$

where $R_a$ is the average roughness, and $R_S$ denotes the average slope of the surface roughness. $L$ is the beam length. It should be mentioned that the higher-order gradients of the surface roughness are neglected [Shaat, 2017].

Utilizing the average parameters defined in equations (19), the equation of motion (equation (16)) and the boundary conditions (equation (18)) can be written in the form:

$$D \frac{\partial^4 w(x,t)}{\partial x^4} + K \frac{\partial^3 w(x,t)}{\partial x^3} + B \frac{\partial^2 w(x,t)}{\partial x^2} - F_z(x,t) + I_0 \ddot{w}(x,t) = 0 \quad (20)$$

$$-D \frac{\partial^3 w(x,t)}{\partial x^3} - \frac{1}{2} K \frac{\partial^2 w(x,t)}{\partial x^2} = \bar{V} \text{ or } w = w^{\Gamma}$$

$$-D \frac{\partial^2 w(x,t)}{\partial x^2} = \bar{M} \text{ or } w_{,x} = w_{,x}^{\Gamma} \quad (21)$$

where

$$D = EI, \text{ i.e. } I = \frac{4}{3} \left( \frac{b}{2} + R_a \right) \left( \frac{h}{2} + R_a \right)^3$$

$$K = \frac{8}{3} E R_S \left[ 3 \left( \frac{b}{2} + R_a \right) \left( \frac{h}{2} + R_a \right)^2 + \left( \frac{h}{2} + R_a \right)^3 \right]$$

$$B = 8 E R_S^2 \left[ \left( \frac{b}{2} + R_a \right) \left( \frac{h}{2} + R_a \right) + \left( \frac{h}{2} + R_a \right)^2 \right] \quad (22)$$

$$I_0 = 2\rho \left( \frac{b}{2} + R_a \right) \left( \frac{h}{2} + R_a \right)$$

It is should be mentioned that for a smooth surface, $R_a \to 0$ and $RS \to 0$. Thus, the model automatically recovers the classical Euler Bernoulli beam model.

## 3. Free vibration of microbeams with rough surfaces

In this section, the derived equation of motion (equation (20)) is solved for the free vibration of cantilever, simple supported, and clamped-clamped beams with rough surfaces. The following nondimensional parameters are employed:

$$W(X) = \frac{\sqrt{12}w(x)}{h}, \quad X = \frac{x}{L}, \quad T = t\sqrt{\frac{D}{I_0 L^4}} \tag{23}$$

where $t$ is the time in seconds while $T$ is the nondimensional time.

According to equation (23), the normalized form of the equation of motion (equation (20)) can be obtained as follows:

$$\frac{\partial^4 W(X,T)}{\partial X^4} + \alpha \frac{\partial^3 W(X,T)}{\partial X^3} + \beta \frac{\partial^2 W(X,T)}{\partial X^2} + \ddot{W}(X,T) = \bar{F}_z(X,T) \tag{24}$$

where

$$\alpha = \frac{KL}{D}$$

$$\beta = \frac{BL^2}{D} \tag{25}$$

$$\bar{F}_z(X,T) = \frac{F_z(x,t)L^4\sqrt{12}}{Dh}$$

and the boundary conditions (equation (21)) can be obtained in the normalized form as follows:

$$\frac{\partial^3 W(X,T)}{\partial X^3} + \frac{1}{2}\alpha \frac{\partial^2 W(X,T)}{\partial X^2} = -\bar{\aleph} \text{ or } W = W^{\Gamma}$$

$$\frac{\partial^2 W(X,T)}{\partial X^2} = -\bar{\mathcal{M}} \text{ or } W_{,X} = W^{\Gamma}_{,X} \tag{26}$$

where $\bar{\aleph}$ and $\bar{\mathcal{M}}$ are the nondimensional applied shear force and bending moment. $W^{\Gamma}$ and $W^{\Gamma}_{,X}$ are the nondimensional prescribed deflection and slope.

The deflection of the beam can be decomposed as follows:

$$W(X,T) = \varphi(X)\exp(i\omega T) \tag{27}$$

where $\omega$ is the nondimensional natural frequency, and $\varphi(X)$ is the corresponding mode shape.

By substituting equation (27) into equation (24) and dropping the forcing term from the result, the equation governing the mode shape of the free vibration of microbeams with rough surfaces is obtained as follows:

$$\frac{d^4\varphi(X)}{dX^4} + \alpha\frac{d^3\varphi(X)}{dX^3} + \beta\frac{d^2\varphi(X)}{dX^2} - \omega^2\varphi(X) = 0 \tag{28}$$

Similarly, the boundary conditions become:

$$\frac{d^3\varphi(X)}{dX^3} + \frac{1}{2}\alpha\frac{d^2\varphi(X)}{dX^2} = 0 \text{ or } \varphi(X) = 0 \tag{29}$$

$$\frac{d^2\varphi(X)}{dX^2} = 0 \text{ or } \frac{d\varphi(X)}{dX} = 0$$

A general solution for equation (28) can be obtained in the form:

$$\varphi(X) = C_1 \exp(\lambda_1 X) + C_2 \exp(\lambda_2 X) + C_3 \exp(\lambda_3 X) + C_4 \exp(\lambda_4 X) \tag{30}$$

where $C_i$ are constants to be determined. $\lambda_i$ are the roots of the following characteristics equation:

$$\lambda^4 + \alpha\lambda^3 + \beta\lambda^2 - \omega^2 = 0 \tag{31}$$

Next, the natural frequencies along with the constants $C_i$ are determined for the different boundary conditions.

### 3.1. Cantilever microbeams

For cantilever micro-beams, the boundary conditions in equation (29) are rewritten as follows:

$$\varphi(0) = 0, \frac{d\varphi(0)}{dX} = 0$$

$$\frac{d^2\varphi(1)}{dX^2} = 0, \frac{d^3\varphi(1)}{dX^3} = 0 \tag{32}$$

To determine the natural frequencies of cantilever microbeams, equation (30) is substituted into equation (32) which gives:

$$\sum_{r=1}^{4} C_r = 0, \sum_{r=1}^{4} C_r \lambda_r = 0, \sum_{r=1}^{4} C_r \lambda_r^2 \exp(\lambda_r) = 0, \sum_{r=1}^{4} C_r \lambda_r^3 \exp(\lambda_r) = 0 \tag{33}$$

Then, the coefficient matrix of the obtained equations is formed. Thus, the nondimensional natural frequencies can be obtained by solving the following equation:

$$\det \begin{bmatrix} 1 & 1 & 1 & 1 \\ \lambda_1 & \lambda_2 & \lambda_3 & \lambda_4 \\ \lambda_1^2 \exp(\lambda_1) & \lambda_2^2 \exp(\lambda_2) & \lambda_3^2 \exp(\lambda_3) & \lambda_4^2 \exp(\lambda_4) \\ \lambda_1^3 \exp(\lambda_1) & \lambda_2^3 \exp(\lambda_2) & \lambda_3^3 \exp(\lambda_3) & \lambda_4^3 \exp(\lambda_4) \end{bmatrix} = 0 \tag{34}$$

The constants, $C_r$, are also determined by solving the first three equations in (equation (33)), as follows:

$$C_3 = -C_1 \frac{\xi_1}{\xi_2}$$

$$C_4 = \frac{1}{\lambda_4 - \lambda_2}(-C_1\lambda_1 + \lambda_2(C_1 + C_3) - C_3\lambda_3)$$

$$C_2 = -C_1 - C_3 - C_4 \tag{35}$$

$$\xi_1 = \lambda_1^2 \exp(\lambda_1) - \lambda_2^2 \exp(\lambda_2) + \left(\frac{\lambda_1 - \lambda_2}{\lambda_4 - \lambda_2}\right)(\lambda_2^2 \exp(\lambda_2) - \lambda_4^2 \exp(\lambda_4))$$

$$\xi_2 = \lambda_3^2 \exp(\lambda_3) - \lambda_2^2 \exp(\lambda_2) + \left(\frac{\lambda_3 - \lambda_2}{\lambda_4 - \lambda_2}\right)(\lambda_2^2 \exp(\lambda_2) - \lambda_4^2 \exp(\lambda_4))$$

### 3.2. Simple supported microbeams

For simple supported microbeams, the boundary conditions (equation (29)) are rewritten as follows:

$$\varphi(0) = \varphi(1) = 0$$
$$\frac{d^2\varphi(0)}{dX^2} = \frac{d^2\varphi(1)}{dX^2} = 0 \tag{36}$$

Similarly, to determine the natural frequencies of simple supported microbeams, equation (36) is substituted into equation (30) which gives:

$$\sum_{r=1}^{4} C_r = 0, \sum_{r=1}^{4} C_r \lambda_r^2 = 0, \sum_{r=1}^{4} C_r \exp(\lambda_r) = 0, \sum_{r=1}^{4} C_r \lambda_r^2 \exp(\lambda_r) = 0 \tag{37}$$

Hence, the nondimensional natural frequencies are determined by solving the following equation:

$$\det \begin{bmatrix} 1 & 1 & 1 & 1 \\ \lambda_1^2 & \lambda_2^2 & \lambda_3^2 & \lambda_4^2 \\ \exp(\lambda_1) & \exp(\lambda_2) & \exp(\lambda_3) & \exp(\lambda_4) \\ \lambda_1^2 \exp(\lambda_1) & \lambda_2^2 \exp(\lambda_2) & \lambda_3^2 \exp(\lambda_3) & \lambda_4^2 \exp(\lambda_4) \end{bmatrix} = 0 \tag{38}$$

Solving the first three equations in (equation (37)) gives the constants, $C_r$, in the form:

$$C_3 = -C_1 \frac{\xi_1}{\xi_2}$$
$$C_4 = \frac{1}{\lambda_4^2 - \lambda_2^2}(-C_1\lambda_1^2 + \lambda_2^2(C_1 + C_3) - C_3\lambda_3^2)$$
$$C_2 = -C_1 - C_3 - C_4 \tag{39}$$
$$\xi_1 = \exp(\lambda_1) - \exp(\lambda_2) + \left(\frac{\lambda_1^2 - \lambda_2^2}{\lambda_4^2 - \lambda_2^2}\right)(\exp(\lambda_2) - \exp(\lambda_4))$$
$$\xi_2 = \exp(\lambda_3) - \exp(\lambda_2) + \left(\frac{\lambda_3^2 - \lambda_2^2}{\lambda_4^2 - \lambda_2^2}\right)(\exp(\lambda_2) - \exp(\lambda_4))$$

### 3.3. Clamped-clamped microbeams

According to equation (29), for clamped-clamped microbeams, the following boundary conditions are employed:

$$\varphi(0) = 0, \varphi(1) = 0, \frac{d\varphi(0)}{dX} = 0, \frac{d\varphi(1)}{dX} = 0 \tag{40}$$

The substitution of equation (40) into equation (30) leads to:

$$\sum_{r=1}^{4} C_r = 0, \sum_{r=1}^{4} C_r \lambda_r = 0, \sum_{r=1}^{4} C_r \exp(\lambda_r) = 0, \sum_{r=1}^{4} C_r \lambda_r \exp(\lambda_r) = 0 \tag{41}$$

Consequently, the nondimensional natural frequencies of clamped-clamped microbeams are obtained by solving the following equation:

$$\det \begin{bmatrix} 1 & 1 & 1 & 1 \\ \lambda_1 & \lambda_2 & \lambda_3 & \lambda_4 \\ \exp(\lambda_1) & \exp(\lambda_2) & \exp(\lambda_3) & \exp(\lambda_4) \\ \lambda_1 \exp(\lambda_1) & \lambda_2 \exp(\lambda_2) & \lambda_3 \exp(\lambda_3) & \lambda_4 \exp(\lambda_4) \end{bmatrix} = 0 \tag{42}$$

The constants, $C_i$, are obtained in the following form by solving the first three equations of (equation (41)):

$$C_3 = -C_1 \frac{\xi_1}{\xi_2}$$

$$C_4 = \frac{1}{\lambda_4 - \lambda_2}(-C_1\lambda_1 + \lambda_2(C_1 + C_3) - C_3\lambda_3)$$

$$C_2 = -C_1 - C_3 - C_4 \qquad (43)$$

$$\xi_1 = \exp(\lambda_1) - \exp(\lambda_2) + \left(\frac{\lambda_1 - \lambda_2}{\lambda_4 - \lambda_2}\right)(\exp(\lambda_2) - \exp(\lambda_4))$$

$$\xi_2 = \exp(\lambda_3) - \exp(\lambda_2) + \left(\frac{\lambda_3 - \lambda_2}{\lambda_4 - \lambda_2}\right)(\exp(\lambda_2) - \exp(\lambda_4))$$

## 4. Effects of surface roughness on frequencies and mode shapes of microbeams

In this section, effects of the surface roughness and the beam size on the natural frequencies of microbeams with different boundary conditions are investigated. Moreover, it is revealed that microbeams exhibit mode localization due to the roughness of their surfaces. To this end, a set of square microbeams ($b = h$) with different thicknesses, $h = [0.05,1]\mu m$, are analyzed. The beam length is considered $L = 50h$. In the performed analyses, results are extracted for microbeams with rough surfaces and compared to those of smooth surfaces. Thus, rough microbeams are considered with an average roughness $R_a = 1\ nm$. To show the influence of surface roughness on the dynamical characteristics of microbeams, results are depicted for different values of the average slope of roughness, $R_S = [0,1]\mu m/L$.

### 4.1. Influence of surface roughness on natural frequencies of microbeams

To demonstrate the significant effects of the surface roughness on the natural frequencies of microbeams, the variations of the first four nondimensional natural frequencies as functions of the average slope of the surface roughness are depicted in Figures 2-4 for different values of the beam thickness and for the different boundary conditions. Moreover, effects of the beam size on the nondimensional natural frequencies of cantilever, simple supported, and clamped-clamped microbeams are depicted in Figure 4. Various nontraditional phenomena can be derived from these figures.

Figure 2 shows effects of the surface roughness on the nondimensional natural frequencies of cantilever beams. As shown in this figure, the first two nondimensional natural frequencies decrease with the increase in the average slope of the surface roughness resulting in a zero-frequency mode. However, the nondimensional natural frequencies of the third mode or the higher modes increase with an increase in the average slope of the surface roughness. An interpretation of these observations is given in Section 4.2.

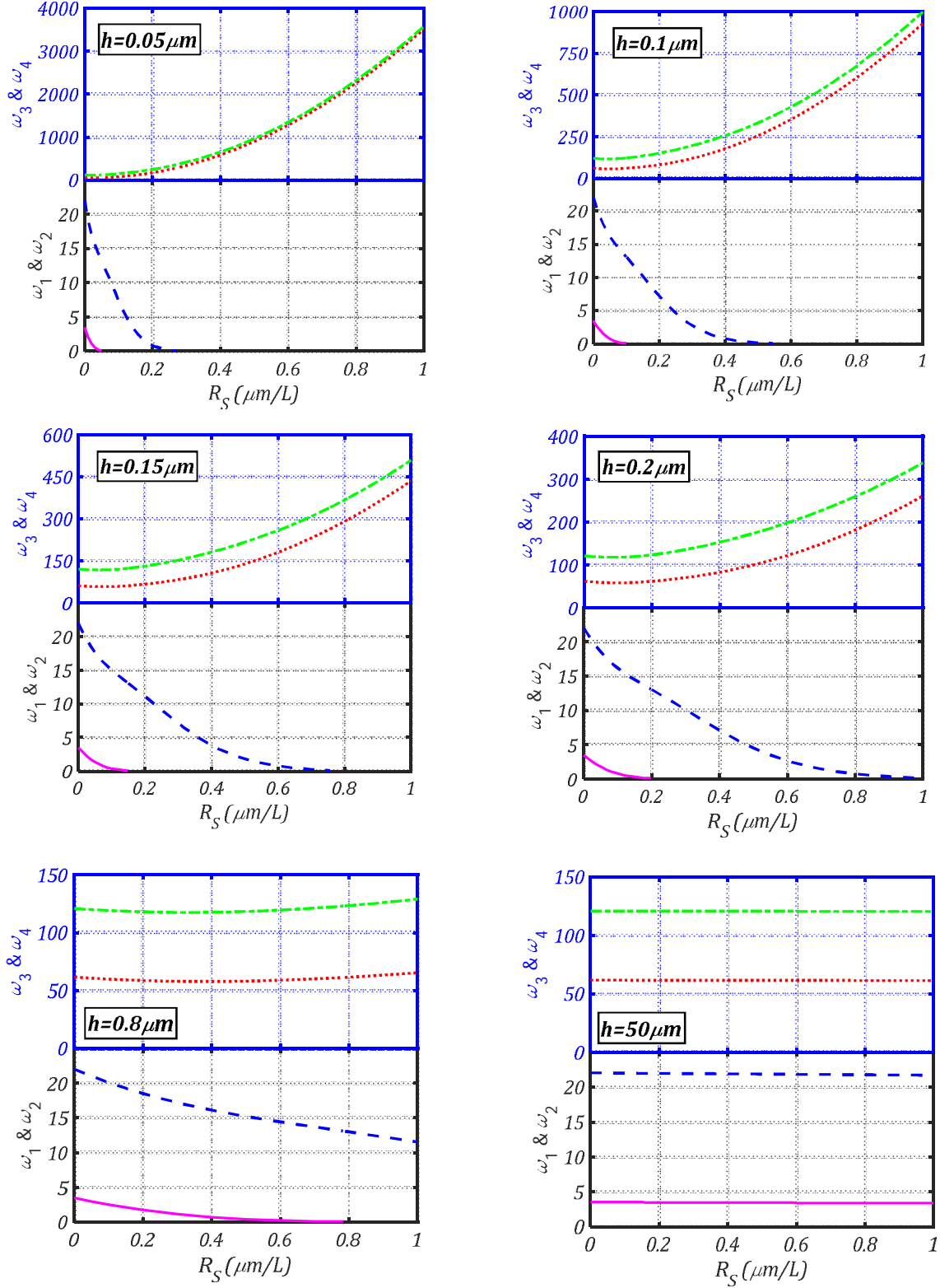

Figure 2: The first four nondimensional natural frequencies as functions of the average slope of the surface roughness, $R_S$, for **cantilever** microbeams ($R_a = 1\,nm$, $L = 50h$, and $b = h$).

Moreover, it follows from Figure 2 that depending on the beam thickness, there is a critical value of the average slope of the surface roughness beyond which the beam starts the oscillation with a high-order mode frequency. For instance, the second mode is the lowest elastic mode of a cantilever microbeam with a thickness $h = 0.15 \mu m$ and an average slope of roughness within the range $0.149 < R_S < 0.75 \mu m/L$. On the other hand, a cantilever microbeam with a thickness $h = 0.15 \mu m$ and an average slope of roughness $R_S > 0.75 \mu m/L$ exhibits a third-order mode oscillation as the initial elastic mode of vibration.

Figure 3 depicts the impact of the surface roughness on the nondimensional natural frequencies of simple supported microbeams. For simple supported microbeams, the fundamental natural frequency decreases with an increase in the average slope of the surface roughness. In contrast, the nondimensional natural frequencies of the higher-order modes increase with an increase in the average slope of the surface roughness. Also, an interpretation of these results is given in Section 4.2.

When compared with cantilever beams, simple supported beams reflect different trends regarding the dependency of their frequencies on the surface roughness. First of all, unlike cantilever microbeams, the nondimensional frequency of the second mode of simple supported microbeams shows an increasing trend as a function of the average slope of roughness. Second, the first two nondimensional natural frequencies of cantilever beams exponentially decay with an increase in the average slope of the surface roughness. On the other hand, the change in the fundamental natural frequency of simple supported beams follows a different trend. This trend is a combination of a power function and an exponentially decaying function. Thus, at small values of the average slope of the surface roughness, the nondimensional fundamental frequency may increase, decrease, or remain constant depending on the beam size, as shown in Figure 3. Then, at larger values of the average slope of roughness, the nondimensional fundamental frequency follows an exponential decaying trend.

The nondimensional natural frequencies of clamped-clamped microbeams are plotted versus the average slope of the surface roughness in Figure 4. Unlike cantilever and simple supported microbeams, all the nondimensional natural frequencies of clamped-clamped microbeams increase with an increase in the average slope of the surface roughness.

It follows from Figures 2-4 and according to the previous discussion, the variations of the natural frequencies of microbeams as functions of the average slope of roughness strongly depend on the boundary conditions. In addition, the inspection of Figures 2-4 leads to the conclusion that the role of the surface roughness of a microbeam increases with a decrease in the beam thickness. Thus, the rate of change in the nondimensional natural frequencies increases with a decrease in the beam thickness. In contrast, the natural frequencies of beams with large thicknesses (*e.g.* $h = 50 \mu m$) do not depend on the surface roughness. At a beam thickness of $h = 50 \mu m$, the natural frequencies of the classical beam (with a smooth surface) are recovered, as shown in Figures 2-4.

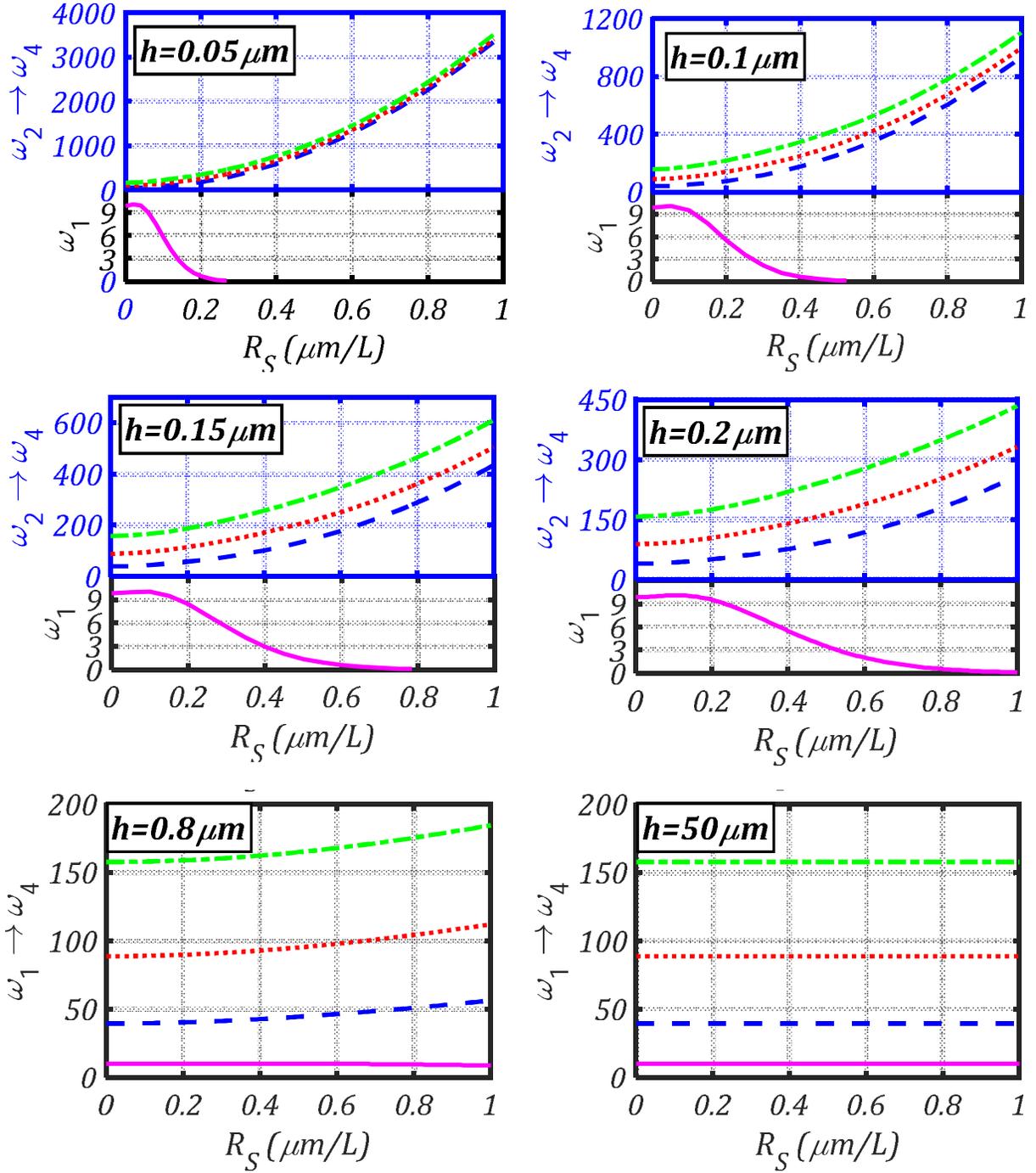

Figure 3: The first four nondimensional natural frequencies as functions of the average slope of the surface roughness, $R_S$, for **simple supported** microbeams ($R_a = 1\ nm$, $L = 50h$, and $b = h$).

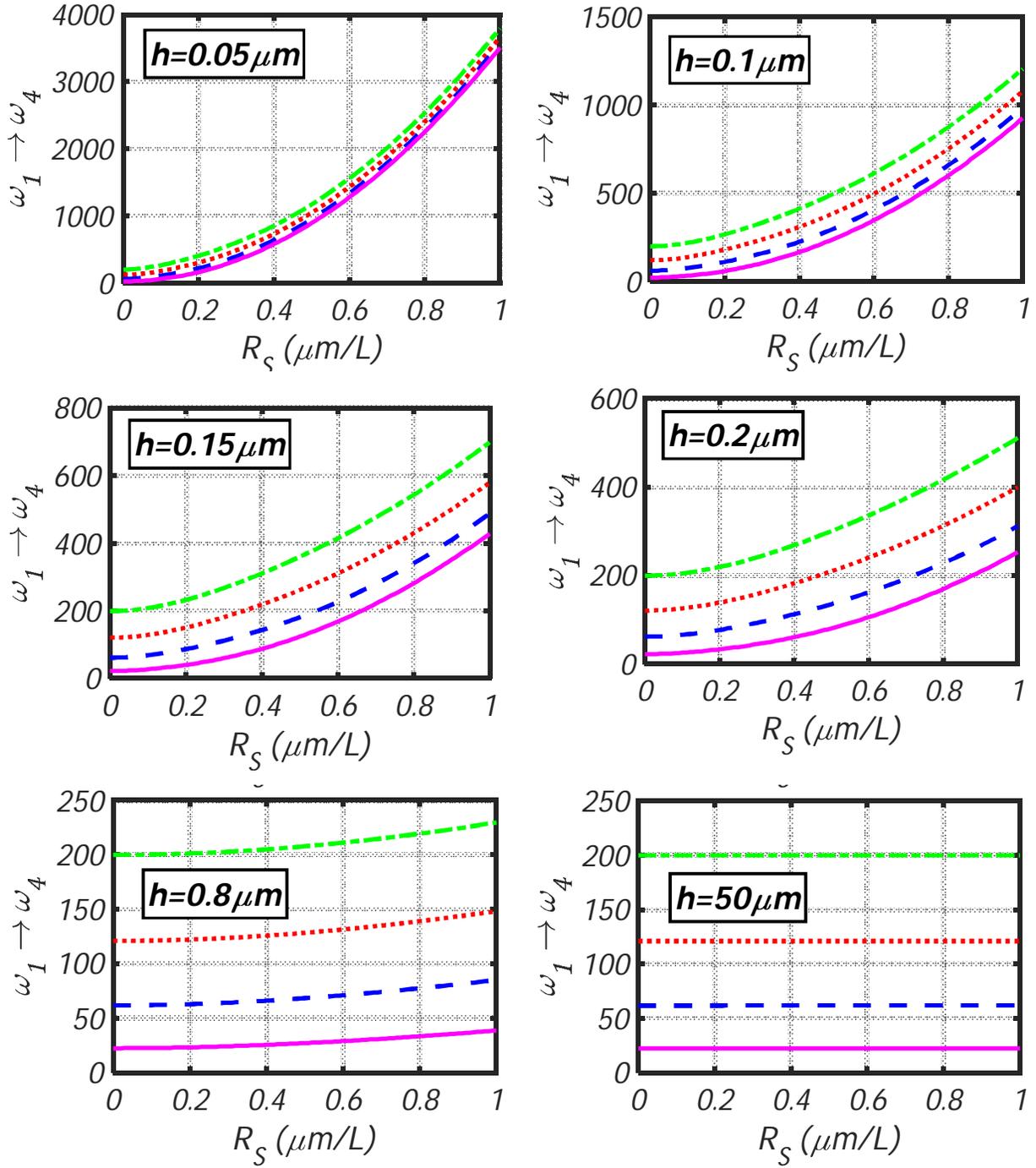

Figure 4: The first four nondimensional natural frequencies as functions of the average slope of the surface roughness, $R_S$, for **clamped-clamped** microbeams ($R_a = 1\ nm$, $L = 50h$, and $b = h$).

Figure 5 depicts the influence of the beam size on the nondimensional natural frequencies of microbeams. As previously demonstrated, a decrease in the beam size is accompanied with an increase in the surface roughness effects. As shown in Figure 5(a), the nondimensional natural frequencies of the first two modes of cantilver beams diminish with a decrease in the beam size. Thus, the fundamental mode and the second mode turn into zero-frequency modes at $h \cong 0.5 \mu m$ and $h \cong 0.1 \mu m$, respectively. As for simple supported microbeams (Figure 5(b)), the fundamental natural frequency converts into a zero-frequency mode at $h \cong 0.1 \mu m$. For all boundary conditions, a nondimensional natural frequency with an increasing trend follows a slow rate enhancement with the decrease in the beam thickness up to $h \cong 0.2 \mu m$. However, a nondimensional natural frequency sharply increases when the beam thickness is reduced lower than $0.2 \mu m$.

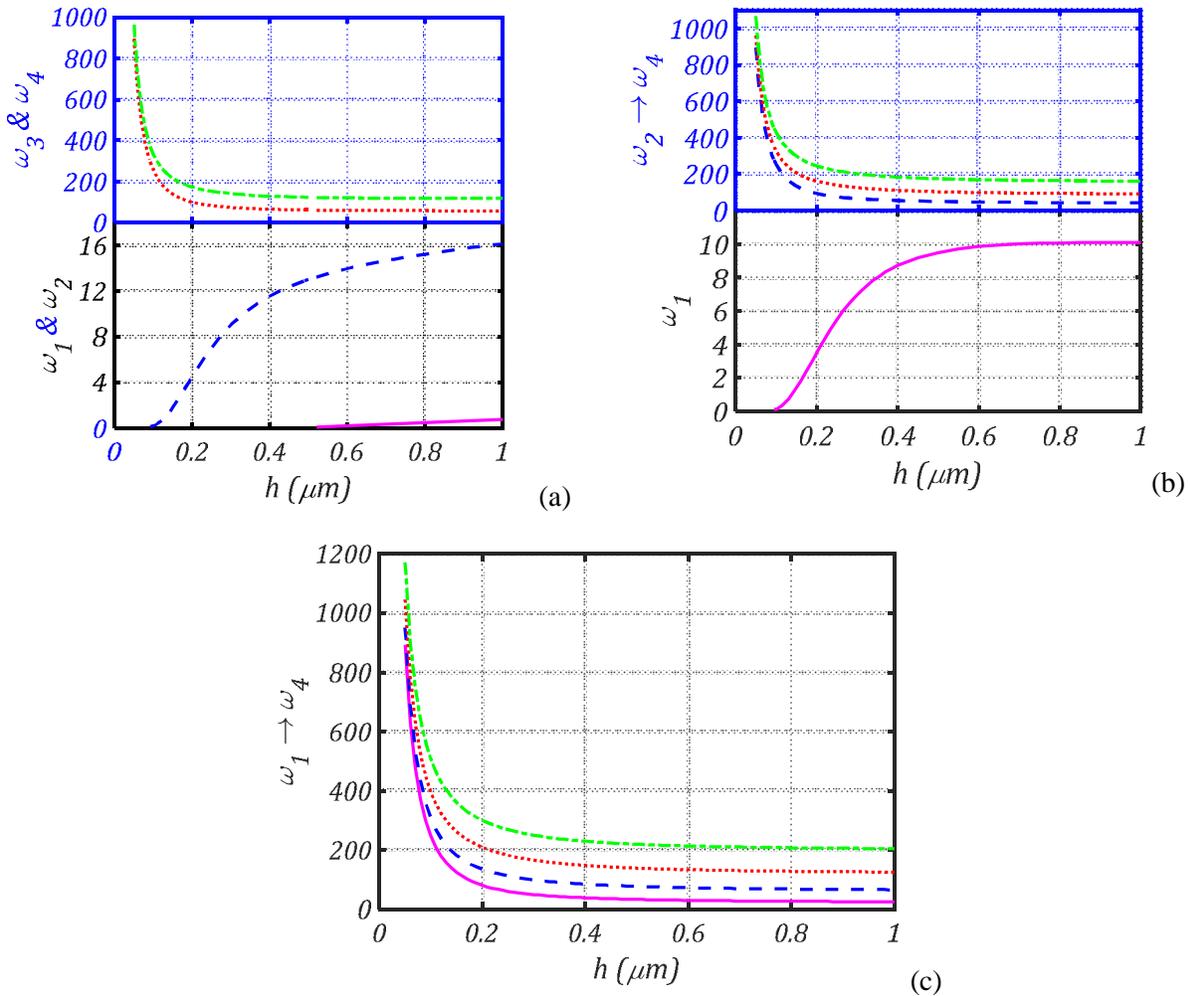

Figure 5: The first four nondimensional natural frequencies as functions of the beam thickness, $h$, for (a) cantilever, (b) simple-supported, and (c) clamped-clamped microbeams ($R_a = 1\ nm$, $R_S = 0.5\ \mu m/L$, $L = 50h$, and $b = h$).

It follows from the depicted results in Figures 2-5 that special considerations should be given for the beam surface roughness when investigating the mechanics of microbeams. Thus, neglecting effects of the surface roughness may result in under/overestimations of the beam natural frequencies.

**4.2. Mode localization phenomenon n microbeams due to surface roughness**

To portray the dependency of the mode shapes of microbeams on the surface roughness, the first four mode shapes of microbeams with different boundary conditions are depicted in Figures 6-8. Figures 6-8 show effects of the surface roughness on modes of vibration of cantilever, simple supported, and clamped-clamped microbeams, respectively. The mode shapes are depicted for different values of the average slope of the surface roughness, $R_S$ showing the evolution of the mode localization with an increase in the average slope of roughness.

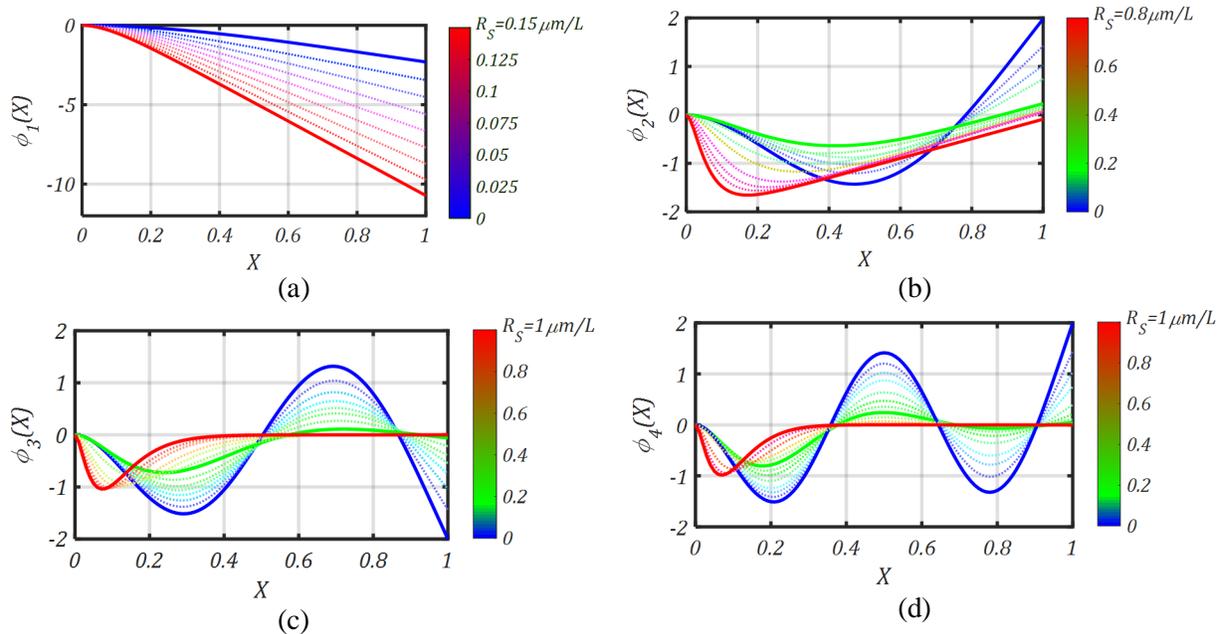

Figure 6: (a) 1st mode shape, (b) 2nd mode shape, (c) 3rd mode shape, and (d) 4th mode shape of **cantilever** microbeams for different values of the average slope of the surface roughness $R_S$ ($h = 0.15 \mu m$, $R_a = 1\ nm$, $L = 50h$, and $b = h$).

Figure 6 shows the mode shapes of a cantilever microbeam with a thickness $h = 0.15 \mu m$. The mode shapes of the considered cantilever beam are obtained with different characteristics depending on the value of the average slope of the surface roughness, $R_S$. The amplitude of vibration of the first mode increases with an increase in the average slope of the surface roughness up to $R_S = 0.149 \mu m/L$, as shown in Figure 6(a). Moreover, the bend of the second mode shape shifts to the fixed end as the average slope of the surface roughness increases up to $R_S = 0.76 \mu m/L$ (Figure 6(b)). For all values of the average slope of roughness, the obtained first two modes are nonlocalized modes where the vibration energy is distributed over the whole beam span, as shown in Figures 6(a) and 6(b). Moreover, for these two modes, the increase in the amplitude indicates that the surface roughness permits more vibration energy to propagate and cover longer spans of the beam. Also, this indicates a softening in the beam stiffness and hence a decrease in the natural frequencies of the first two modes; the result which was demonstrated in Figure 2.

Figures 6(c) and 6(d) show that the 3rd and 4th modes of vibration can be localized due to surface roughness. Thus, the surface roughness inhibits the propagation of the vibration energy throughout the beam length. The inhibition of the vibration energy increases with an increase in the average slope of the surface roughness. As a result, the magnitude of vibration decreases, and the vibration energy is confined within a small portion of the beam leading to a mode localization. The decrease in the amplitude of vibration is an indication of the increase in the beam natural frequency, as previously revealed in Figure 2. Moreover, the increase in the average slope of the surface roughness converts the mode from being weakly localized into a strong mode localization. For instance, a microbeam with an average surface roughness slope $R_S = 0.06 \mu m/L$ is weakly localized where the amplitude of vibration near the fixed end is much higher than the amplitude of vibration of the other portion of the beam. The length of the portion of the beam over which the vibration energy is confined decreases with an increase in the average slope of the surface roughness. Thus, when $R_S = 1 \mu m/L$, the 3rd mode and 4th mode are strongly localized where the vibration energy is, respectively, confined near the fixed end over 43% and 34% of the beam length, as shown in Figures 6(c) and 6(d).

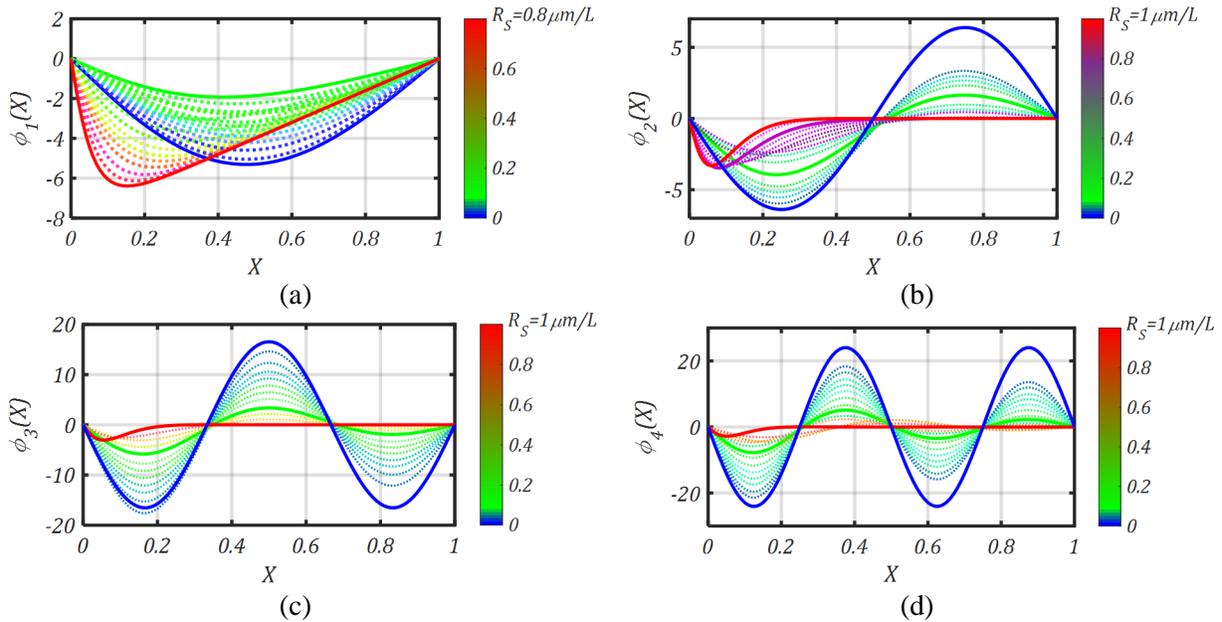

Figure 7: (a) 1st mode shape, (b) 2nd mode shape, (c) 3rd mode shape, and (d) 4th mode shape of **simple supported** microbeams for different values of the average slope of the surface roughness $R_S$ ($h = 0.15 \mu m$, $R_a = 1\ nm$, $L = 50h$, and $b = h$).

Figure 7 reflects the significant effects of the surface roughness on the first four mode shapes of a simple supported microbeam with a thickness $h = 0.15 \mu m$. As presented in Figure 7(a), no mode localization experienced by the simple supported beam for the fundamental mode shape where the vibration energy is distributed over the whole beam length. The magnitude of the fundamental mode decreases with an increase in the average slope of the surface roughness up to $R_S \cong 0.08 \mu m/L$ indicating an inhibition of the vibration energy due to the surface roughness. The latter observation explains the small increase in the nondimensional natural frequency observed in Figure 3 at small values of the average slope of roughness. Beyond the aforementioned value of $R_S$, the amplitude of vibration increases with an increase in the average slope of the surface roughness indicating a decrease in the fundamental natural frequency of the simple supported beam (see Figure 3).

As for the higher-order modes of simple supported beams, the evolution of the mode localization with the increase in the average slope of the surface roughness can be observed in Figures 7(b)-7(d). An increase in the average slope of the surface roughness is accompanied with an increase in the inhibition of vibration and an increase in the vibration confinement. As a consequence, higher-order modes strongly localized at high values of the average slope of the surface roughness. Moreover, the localization of the mode and the inhibition of the vibration energy increases with the increase in the mode number. At $R_S = 1 \mu m/L$, the vibration energy of the 2nd, 3rd, and 4th modes is, respectively, confined over 34%, 22%, and 19% of the beam length. Also, it can be observed that an increase in the average slope of the surface roughness is accompanied with a decrease in the amplitude of vibration indicating an increase in the natural frequency of the corresponding mode. The latter observation explains the increase in the nondimensional natural frequencies of the higher-order modes of simple supported beams due to surface roughness (see Figure 3).

The evolution of the mode localization of a clamped-clamped microbeam with a thickness $h = 0.15 \mu m$ is presented in Figure 8. Unlike cantilever and simple supported beams, all the modes of the clamped-clamped microbeam are localized at high values of the average slope of the surface roughness. Thus, when the average slope of the surface roughness is $R_S = 1 \mu m/L$, the vibration energy of the first four modes is, respectively, confined within 45%, 38%, 30%, and 25% of the beam length, as shown in Figure 8. Moreover, the amplitude of vibration decreases with an increase in the average slope of the surface roughness explaining the increase in the natural frequencies of clamped-clamped microbeams observed in Figure 4.

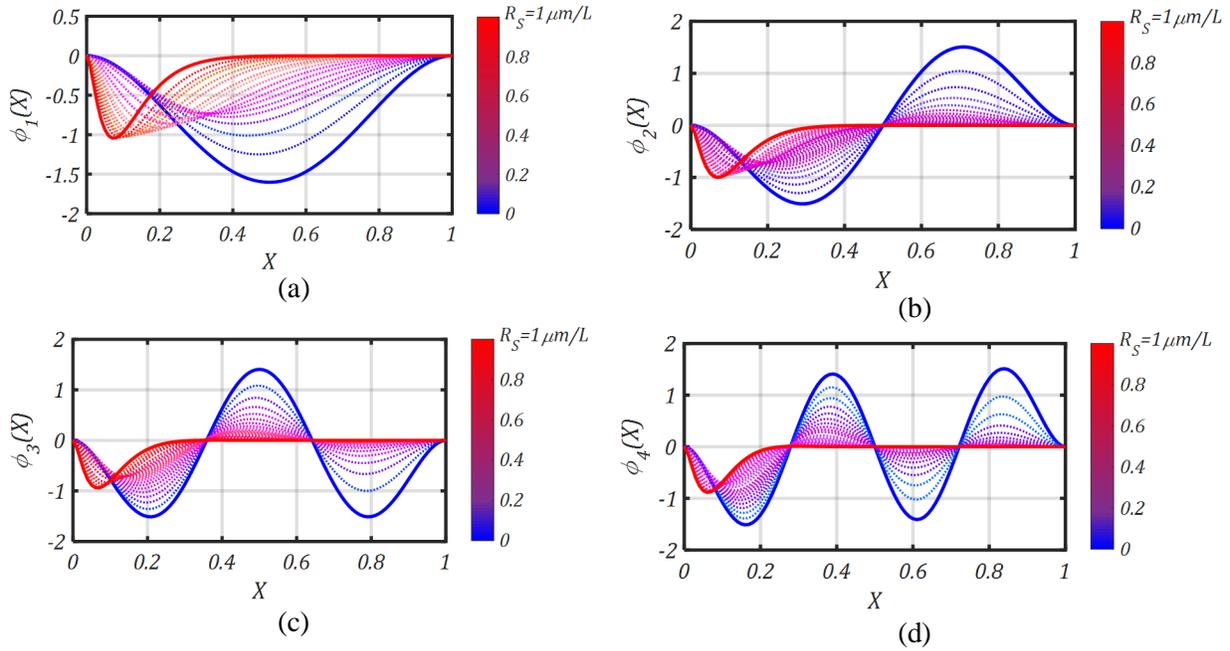

Figure 8: (a) 1st mode shape, (b) 2nd mode shape, (c) 3rd mode shape, and (d) 4th mode shape of **clamped-clamped** microbeams for different values of the average slope of the surface roughness $R_S$ ($h = 0.15 \mu m$, $R_a = 1\ nm$, $L = 50h$, and $b = h$).

Generally speaking, it follows from the presented results in Figures 2-8 that a mode localization is conjugate to a natural frequency increase. However, when a frequency decreases, no mode localization takes place due to surface roughness.

## Conclusions

This paper presented the first study on the mode localization phenomenon in microbeams due to surface roughness. The equation of the motion of microbeams with rough surfaces was derived incorporating measures for the average surface roughness and the average slope of the surface asperities. Moreover, the derived model was solved for the free vibration of cantilever, simple supported, and clamped-clamped beams. The variations of the natural frequencies of microbeams with rough surfaces were reported as functions of the average slope of the surface roughness and the beam size. Moreover, investigations on the evolution of the mode localization with an increase in the average slope of roughness were carried out.

The presented results came demonstrating two prospects. It was revealed that surface roughness may lead to a zero-frequency mode or a mode localization. As for the first prospect, surface roughness may give more freedom for vibration energy to propagate through the beam span, and hence the mode shapes are nonlocalized. Thus, the surface roughness add more softness to the corresponding mode of the beam and reduces its natural frequency. It was demonstrated that the first two natural frequencies of cantilever microbeams and the fundamental frequency of simple supported beams decrease due to surface roughness. Thus, at high values of the average slope of the surface roughness, these modes of vibration convert into zero-frequency modes where frequencies attain zeros. As for the second prospect, surface roughness may inhibit the propagation of vibration energy throughout the beam length leading to a mode localization. Thus, the vibration energy is confined over a small portion of the beam causing an increase in the natural frequency. The higher-order modes of vibration of cantilever and simple supported beams exhibit strong mode localization at high values of the average slope of the surface roughness. As for clamped-clamped beams, all modes of vibration are localized at high values of the average slope of the surface roughness. Generally, the mode localization is accompanied with an increase in the natural frequency of the microbeam.